\documentclass[journal]{IEEEtran}
 \pdfoutput=1
 \usepackage{cite}
\usepackage{amsmath,amssymb,amsfonts}
\usepackage{algorithmic}
\usepackage{graphicx}
\usepackage{textcomp}
\usepackage{framed,multirow}
\usepackage{latexsym}
\usepackage{amsmath}
\usepackage{threeparttable}
\usepackage{float}
\usepackage{array}
\usepackage{hhline}
\usepackage{colortbl}
\usepackage{stfloats}
\usepackage{booktabs}
\usepackage{multirow}
\usepackage{enumerate}
\usepackage{diagbox}
\usepackage{url}
\usepackage{hyperref}

\def\etal{\textit{et~al}.}

\DeclareGraphicsExtensions{.pdf,.jpeg,.png,.jpg}
\DeclareMathOperator*{\argmax}{argmax}
\graphicspath{{fig/},{samples/},{samples/BCSS/}}

\definecolor{mygray}{gray}{.9}

\definecolor{newcolor}{rgb}{.8,.349,.1}

\makeatletter

\newcommand{\Rmnum}[1]{\expandafter\@slowromancap\romannumeral #1@}
\makeatother

\title{Multi-Layer Pseudo-Supervision for Histopathology Tissue Semantic Segmentation using Patch-level Classification Labels}

\author{Chu Han$^\dagger$, \IEEEmembership{Member, IEEE}, Jiatai Lin$^\dagger$, Jinhai Mai$^\dagger$, Yi Wang, Qingling Zhang, Bingchao Zhao, Xin Chen, Xipeng Pan, Zhenwei Shi, Xiaowei Xu, Su Yao, Lixu Yan, Huan Lin, Zeyan Xu, Xiaomei Huang, Guoqiang Han, Changhong Liang, Zaiyi Liu
\thanks{This project is supported by the Key R\&D Program of Guangdong Province (No. 2021B0101420006), the National Science Fund for Distinguished Young Scholars (No. 81925023), the National Natural Science Foundation of China (No. 62102103, 82102034, 62002082, 82072090, 81771912 and 82071892), The Guangdong Natural Science Foundation (No. 2017A030312008).}
\thanks{Chu Han, Xipeng Pan and Zhenwei Shi are with the Department of Radiology, Guangdong Provincial People's Hospital, Guangdong Academy of Medical Sciences, Guangzhou, China; 
Guangdong Cardiovascular Institute, Guangzhou, China. E-mail: hanchu@gdph.org.cn}
\thanks{Jinhai Mai, Bingchao Zhao, Huan Lin, Zeyan Xu, Xiaomei Huang, Changhong Liang and Zaiyi Liu are with the Department of Radiology, Guangdong Provincial People’s Hospital, Guangdong Academy of Medical Sciences, Guangzhou, Guangdong, 510080, China.}
\thanks{Jiatai Lin and Guoqiang Han are with the School of Computer Science and Engineering, South China University of Technology, Guangzhou, China}
\thanks{Xiaowei Xu is with Guangdong Cardiovascular Institute, Guangdong Provincial Key Laboratory of South China Structural Heart Disease, Guangdong Provincial People's Hospital, Guangdong Academy of Medical Sciences, Guangzhou, 510080, China}
\thanks{Qingling Zhang, Su Yao and Lixu Yan are with the Department of Pathology, Guangdong Provincial People's Hospital, Guangdong Academy of Medical Sciences, Guangzhou, Guangdong, 510080, China}
\thanks{Yi Wang is with the National-Regional Key Technology Engineering Laboratory for Medical Ultrasound, Guangdong Key Laboratory for Biomedical Measurements and Ultrasound Imaging, School of Biomedical Engineering, Health Science Center, Shenzhen University, Shenzhen, China}
\thanks{Xin Chen is with the Department of Radiology, Guangzhou First People's Hospital, the Second Affiliated Hospital of South China University of Technology, Guangzhou, 510180, China}
\thanks{Corresponding author: Guoqiang Han, Changhong Liang, Zaiyi Liu.}
\thanks{$^\dagger$The first three authors contribute equally. }
}
\begin{document}
\maketitle

\IEEEtitleabstractindextext{\begin{abstract}
Tissue-level semantic segmentation is a vital step in computational pathology. Fully-supervised models have already achieved outstanding performance with dense pixel-level annotations. However, drawing such labels on the giga-pixel whole slide images is extremely expensive and time-consuming. In this paper, we use only patch-level classification labels to achieve tissue semantic segmentation on histopathology images, finally reducing the annotation efforts. We proposed a two-step model including a classification and a segmentation phases. In the classification phase, we proposed a CAM-based model to generate pseudo masks by patch-level labels. In the segmentation phase, we achieved tissue semantic segmentation by our proposed Multi-Layer Pseudo-Supervision. Several technical novelties have been proposed to reduce the information gap between pixel-level and patch-level annotations. As a part of this paper, we introduced a new weakly-supervised semantic segmentation (WSSS) dataset for lung adenocarcinoma (LUAD-HistoSeg). We conducted several experiments to evaluate our proposed model on two datasets. Our proposed model outperforms two state-of-the-art WSSS approaches. Note that we can achieve comparable quantitative and qualitative results with the fully-supervised model, with only around a 2\% gap for MIoU and FwIoU. By comparing with manual labeling, our model can greatly save the annotation time from hours to minutes. The source code is available at: \url{https://github.com/ChuHan89/WSSS-Tissue}.
\end{abstract}
}

\maketitle
\IEEEdisplaynontitleabstractindextext

\IEEEpeerreviewmaketitle

\section{Introduction}
\begin{figure}[t]
	\centering
	\setlength{\tabcolsep}{1pt}
	\begin{tabular}{cccc}
		\includegraphics[width=.245\linewidth]{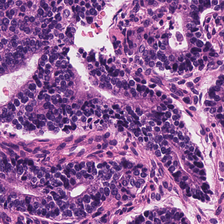}&
		\includegraphics[width=.245\linewidth]{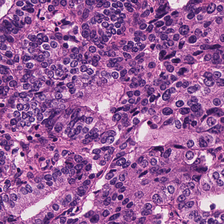}&
		\includegraphics[width=.245\linewidth]{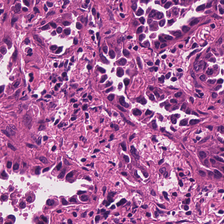}&
		\includegraphics[width=.245\linewidth]{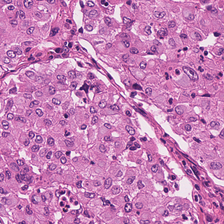}\\
	\end{tabular}
	\caption{Demonstration of tumor heterogeneity. Four patches with tumor epithelial from the WSIs of four lung adenocarcinoma patients.}
	\label{fig:tumor-heter}
\end{figure}
Tumor microenvironment (TME), not only plays a vital role in tumor initiation and progression~\cite{hanahan2011hallmarks}, but also influences the therapeutic effect and prognosis of cancer patients~\cite{skrede2020deep,abduljabbar2020geospatial}. TME is formed with different types of tissues, including tumor epithelial, tumor-infiltrating lymphocytes (TILs), tumor-associated stroma and etc. They have been proven to be clinically relevant with tumor progression by previous studies. TILs was considered as prognostic biomarkers in numerous solid tumors, such as lung cancer~\cite{brambilla2016prognostic}, breast cancer~\cite{denkert2018tumour} and colorectal cancer~\cite{kong2019prognostic}. While the crosstalk between tumor epithelial and tumor-associated stroma has been associated with tumor progression~\cite{mao2013stromal,bremnes2011role}.
Therefore, it is essential to differentiate and segment different types of tissues for precise quantification of TME.

Conventional approaches perform tissue segmentation by using hand-crafted features, such as textures~\cite{diamond2004use,sirinukunwattana2015novel}, morphological features~\cite{anoraganingrum1999cell}, color~\cite{tabesh2007multifeature} and etc. Recently, deep learning~\cite{lecun2015deep} demonstrates its superiority and shows tremendous success in medical image segmentation tasks~\cite{chen2016dcan,zhao2020nuclei,qaiser2019fast}. However, collecting dense pixel-level annotations is expensive and labor-intensive, especially for histopathology images. Because of the diversity and complexity, only pathologists or people with the clinical background can handle it. Moreover, due to the heterogeneity and the aggressiveness of tumors, it may have various morphological appearances, as demonstrated in Fig~\ref{fig:tumor-heter}. In the meanwhile, the gigapixel of the whole slide image (WSI) also increases the difficulty of manual labeling.

Researchers have made attempts to overcome the difficulties of dense annotation acquisition and reduce the annotation efforts, such as active learning~\cite{yang2017suggestive}, semi-supervised learning~\cite{liu2020semi}, learning from sparse annotations~\cite{liang2018weakly}, and weakly-supervised learning~\cite{gao2020renal}. Class activation mapping~\cite{zhou2016learning} is the most common solution for weakly-supervised semantic segmentation (WSSS). The basic idea is to train a classification network and leverage the semantic information from the deeper layers to achieve segmentation. Chan~\etal~\cite{chan2019histosegnet} applied a CAM-based method with a series of post-processing steps for weakly-supervised tissue semantic segmentation. However, CAM-based methods face a great challenge that the classification network tends to differentiate objects by the most discriminative features but the segmentation task aims to find the complete object. The activated regions will gradually shrink and harm the segmentation results. For histopathology images, such contradiction will be amplified because the spatial arrangement of different tissues is relatively random and scatter comparing with the natural images.

In this paper, we present a simple and effective CNN model for histopathology tissue semantic segmentation using only patch-level annotations. Pathologists only need to judge the presence or absence of the different tissue categories in the patches instead of carefully drawing the boundaries of tissues on the WSIs, which greatly saves the annotation time. The basic idea of this model is to use patch-level classification labels to automatically generate pixel-level semantic segmentation masks, and then use the generated pseudo masks to train a semantic segmentation model. 

Our proposed model contains a classification phase and a segmentation phase. In the classification phase, we proposed a CAM-based classification model for pseudo mask generation. To avoid the discriminative region shrinkage problem, we proposed a Progressive-Dropout Attention (PDA) to progressively deactivate the highlighted regions, and push the classification network to differentiate the tissue categories by the non-predominant regions. In the segmentation phase, we train a semantic segmentation model by the pseudo masks generated from multiple layers of the classification network, we called it Multi-Layer Pseudo-Supervision (MLPS). MLPS can provide information from different stages to reduce the information gap between patch-level and pixel-level labels. Due to the long-tail and unbalanced distribution problem, some tissue categories with fewer training samples may not be able to learn a good feature representation from pseudo masks, which could easily lead to false-positive segmentation results. To tackle this problem, we proposed a classification gate mechanism to reduce the false-positive rate for the non-predominant tissue categories.

In addition, we introduced a new weakly-supervised tissue semantic segmentation dataset for lung adenocarcinoma (LUAD-HistoSeg), which is the first tissue-level semantic segmentation dataset for LUAD. There are four different types in this dataset, tumor epithelial (TE), tumor-associated stroma (TAS), lymphocyte (LYM) and necrosis (NEC), including 16,678 patches with one-hot encoding labels and 607 patches with pixel-level labels under 10$\times$ magnification. 

We evaluate our proposed model on two datasets, LUAD-HistoSeg and Breast Cancer Semantic Segmentation (BCSS)~\cite{BCSS}. Extensive experiments and ablation studies have demonstrated the superiority of our proposed model on semantic segmentation using only patch-level annotations. Comparing with the fully-supervised model, our model shows comparable quantitative and qualitative results with only around a 2\% gap for MIoU and FwIoU. The proposed model has been proven to be $10\times$ faster than manual labeling. The main contributions of this paper are summarized as follows:
\begin{itemize}
	\item We present a tissue semantic segmentation model for histopathology images using only patch-level classification labels, which greatly saves the annotation time for pathologists.
	\item Multi-layer pseudo-supervision with progressive dropout attention is proposed to reduce the information gap between patch-level and pixel-level labels. And a classification gate mechanism is introduced to reduce the false-positive rate.
	\item Our proposed model achieves state-of-the-art performance comparing with weakly-supervised semantic segmentation models on two datasets, as well as a comparable performance with fully-supervised baseline.
	\item The first LUAD dataset is released for weakly-supervised tissue semantic segmentation.
\end{itemize} 

\section{Related Works}
\subsection{Histopathology Image Segmentation}
Computational pathology~\cite{srinidhi2020deep,2020Deep} has attracted much attention in recent years with the advance of deep learning techniques. Histopathology image segmentation is the most vital process in computer-aided histopathology image analysis. With the data-driven nature, various segmentation approaches have been carried out and achieved outstanding performance, such as tissue segmentation~\cite{HookNet}, gland segmentation~\cite{chen2016dcan, 9164951}, nuclei segmentation~\cite{zhao2020nuclei,graham2019hover} and etc.

To prepare sufficient manually labeled data for training CNN models, pathologists have to carefully draw pixel-level labels on a gigapixel whole slide image, which is extremely expensive and time-consuming. Due to the heterogeneity of malignant tumors, even the same tumor type can show totally different morphological appearances. Therefore, people without clinical backgrounds are not qualified for this job. How to reduce the annotation efforts is still an open problem.

\subsection{Reducing Annotation Efforts for Medical Image Segmentation}
Recently, researchers attempt to reduce the costly annotation burden in different technical perspectives, such as active learning, coarse segmentation by patch-level classification, semi-supervised learning and weakly-supervised learning.

\subsubsection{Active Learning}
Active learning (AL)~\cite{settles2009active,budd2021survey} is a human-in-the-loop learning strategy for alleviating annotation burden. It allows human annotators to revisit and refine the uncertain pseudo-labels generated by machine learning algorithms~\cite{wen2018comparison}, or automatically selects the most informative samples to be labeled next~\cite{yang2017suggestive,zhou2021active}. Mahapatra~\etal~\cite{mahapatra2018efficient} proposed to associate GAN with AL to further reduce the annotation effort which saves around 65\% annotations for classification and segmentation on a chest XRay dataset. Doyle~\etal~\cite{Doyle2011} proposed AL with a class-balancing strategy to solve the minority class problem, which achieved a higher accuracy for patch-level classification of non-cancer regions in prostate histopathology image comparing with random sampling strategy. Belharbi~\etal~\cite{belharbi2021deep} proposed to incorporate classification with image-level annotations and segmentation with generated pseudo pixel-level labels for both histopathology image segmentation (gland segmentation) and natural image segmentation (bird species). A deeply supervised active learning (DSAL)~\cite{zhao2021dsal} has been proposed to assign the samples with high uncertainties to strong labelers and the samples with low uncertainties to weak labelers. Shen~\etal~\cite{shen2020deep} considered dissatisfaction, representativeness and diverseness of the samples in AL sampling strategy, for breast cancer segmentation in IHC whole slide images.

\subsubsection{Patch-level Classification}
An alternative solution to alleviate dense pixel-level annotations is to reform the semantic segmentation problem to the patch-level classification problem. And a series of studies have proven the effectiveness of this way by successful diagnostic and survival prediction~\cite{Kather2019,kather2019predicting,hou2016patch}. Several patch-level histopathology classification models have been proposed to avoid densely pixel-level annotation. The most common way is to transfer a classification model trained on ImageNet~\cite{krizhevsky2012imagenet} to the target histopathology domain. Ni~\etal~\cite{ni2019wsi} reduced the inference speed by discarding easier-recognized non-malignant regions in the lower layers and let the higher layer focus on differentiating more complex cancerous regions. Rkaczkowski~\etal~\cite{rkaczkowski2019ara} proposed an accurate, reliable and active model for histopathology image classification, which segmented whole slide images of colorectal cancer into eight tissue types. However, patch-level classification can only perform rough segmentation and sacrifices the pixel-level accuracy for a computationally efficient inference time and lower annotation efforts.

\subsubsection{Semi-supervised Learning}
Semi-supervised learning~\cite{blum1998combining} aims to leverage a small set of labeled samples and a large set of unlabeled samples to train the model. To maximize the value of the limited labels, existing works either try to maintain the consistency by competing for the introduced perturbations~\cite{laine2016temporal,10.5555/3294771.3294885} or seek the relationship among different samples~\cite{liu2019knowledge,battaglia2018relational}. Self-supervised learning~\cite{zhai2019s4l,cheng2020selfsimilarity,cai2021exponential,kim2021selfmatch} is a feasible way to learn the visual representation for semi-supervised learning, which can somehow be a complement to the lack of annotations. Specific to medical image segmentation, Xia~\etal~\cite{xia2020uncertainty} proposed uncertainty-aware multi-view co-training for 3D volumetric medical image segmentation. Marini~\etal~\cite{marini2021semi} used a semi-supervised semantic segmentation teacher model to train a semi-weakly supervised student model and achieved prostate histopathology image classification. Li~\etal~\cite{li2020self} proposed self-loop uncertainty to generate pseudo labels by a Jigsaw puzzle-solving self-supervised task. Xie~\etal~\cite{xie2020pairwise} introduced a pair relation network (PR-net) to learn a better image representation by comparing a pair of images in the feature space. Then the well-trained PR-net could be transferred to a gland segmentation network in a semi-supervised learning manner.

\subsubsection{Weakly-supervised Learning}
Besides scarcer annotations with semi-supervised learning, many works focus on using weaker or sparser labels for model training, such as image-level labels~\cite{wang2020self}, point-based annotations~\cite{bearman2016s}, bounding boxes~\cite{dai2015boxsup}, scribbles~\cite{lin2016scribblesup} and etc.

Researchers proposed multiple instance learning models with weak supervision for cancerous region segmentation~\cite{jia2017constrained,lerousseau2020weakly}. Lee~\etal~\cite{lee2020scribble2label} proposed to use scribble annotations to automatically generate pseudo-labels for microscopic cell segmentation. Qu~\etal~\cite{qu2019weakly,qu2020weakly} proposed a two-stage weakly supervised learning model with only a small set of point annotations for nuclei segmentation. They first used a semi-supervised model to detect the center points of all the nuclei, and then designed a weakly supervised model for nuclei segmentation. Tokunaga~\etal~\cite{Pseudo_Labeling} leveraged the proportion of the tissue subtypes to generate pseudo labels. Zhang~\etal~\cite{ZHANG2021102183} used foreground proportion as the weak labels and then combine FCN and graph convolutional networks (FGNet) for automatic tissue segmentation.
Wang~\etal~\cite{wang2019weakly} proposed a ScanNet to first train a classification model under the lower resolution and inference with an FCN structure under the higher resolution. With the predicted heatmap of cancerous regions, they can differentiate different types of lung carcinomas. To predict Gleason grades of prostate cancer, Silva-Rodr{\'\i}guez~\etal~\cite{silva2021weglenet} proposed a weakly supervised semantic segmentation (WSSS) model to distinguish morphological appearances of different Gleason grades at the local-level. Inspired by the researches on class activation maps~\cite{zhou2016learning}, Chan~\etal~\cite{chan2019histosegnet} proposed a CAM-based WSSS model with a series of post-processing steps.

In this paper, we aim to achieve tissue-level semantic segmentation by using only patch-level labels.
\begin{figure*}[t]
	\centering
	\includegraphics[width=.99\linewidth]{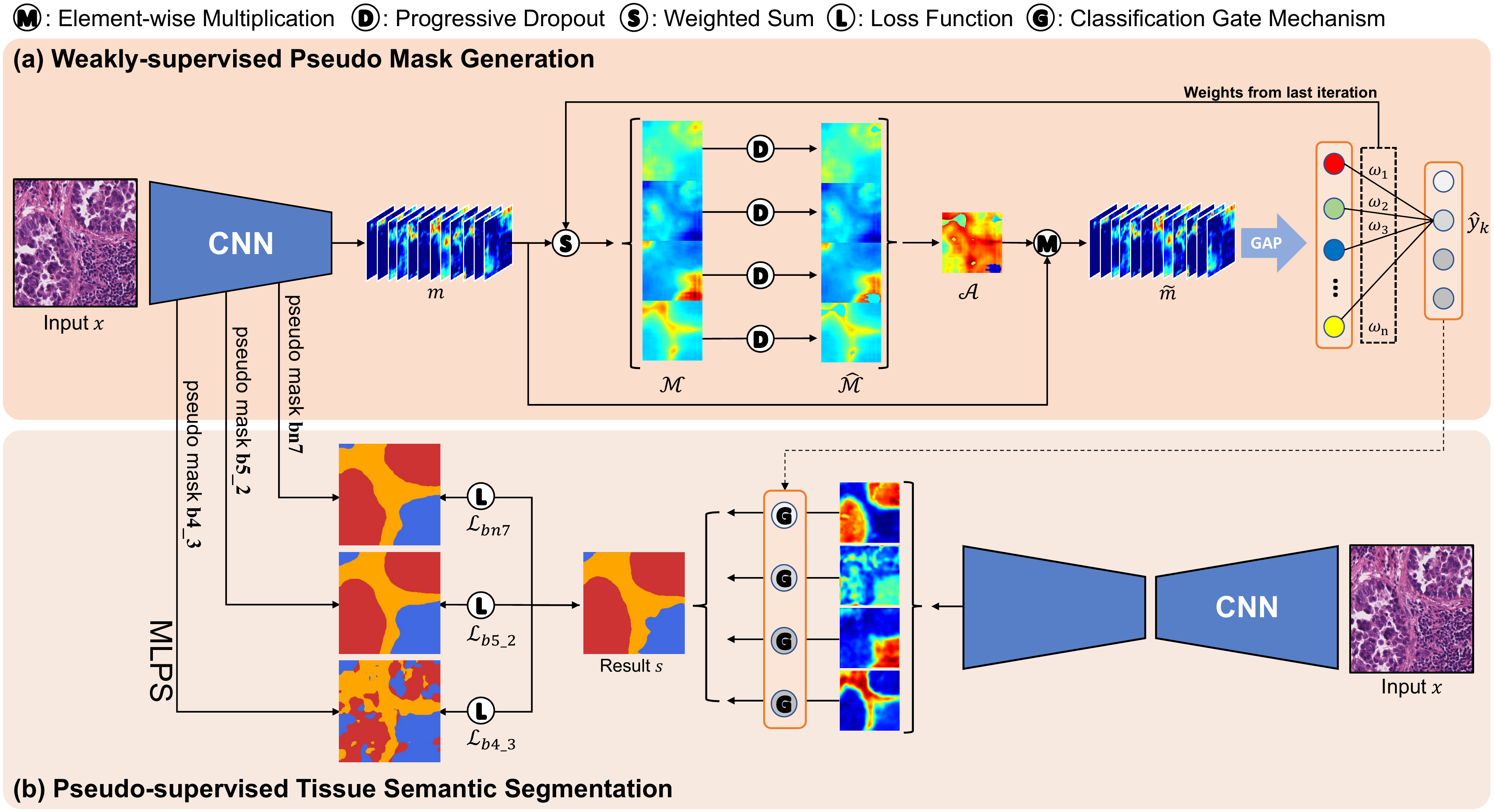}
	\caption{Weakly supervised tissue semantic segmentation architecture. (a) A weakly-supervised model with Progressive Dropout Attention (PDA) was proposed to generate multi-layer pseudo masks for tissue semantic segmentation. (b) DeepLab V3+ model with a proposed classification gate mechanism was introduced for semantic segmentation, guided by Multi-Layer Pseudo Supervision (MLPS). Theoretically, this model can support semantic segmentation for any number of tissue categories. In this figure, we demonstrate the model using the patch example from our proposed dataset LUAD-HistoSeg with four tissue classes (tumor epithelial, tumor-associated stroma, necrosis and lymphocyte).}
	\label{fig:network}
\end{figure*}

\section{Methodology}
Manual labeling dense pixel-level annotations for histopathology images is extremely difficult and time-consuming. In this paper, we proposed a tissue semantic segmentation model by using only patch-level labels in order to alleviate the annotation efforts. Fig.~\ref{fig:network} demonstrates the systematic design of our proposed model. In Section~\ref{sec:phase1}, we trained a patch-level multi-label classification network with our proposed progressive dropout attention to generate pixel-level pseudo masks. In Section~\ref{sec:phase2}, we proposed multi-layer pseudo-supervision to train the semantic segmentation model. A classification gate mechanism is proposed to further guide the segmentation results to reduce the false positive rate.

\textbf{Phase One (classification):} Let us denote the given training data in the classification phase as $\mathcal{D}_{cls}= \{(x, y)| x\in \mathcal{X}, y \in \mathcal{Y}\}$, where $x$ is the patch cropped from the whole slide images and $y$ is the one-hot encoding vector representing the presence or absence of every tissue category in $x$. $\hat{y}=f_{cls}(x,\phi_{cls})$ is the multi-label predicted classification result by the classification model $f_{cls}$ with parameters $\phi_{cls}$. The objective of this phase is to use only the patch-level label $y$ to generate the dense pixel-level pseudo mask $p$.
\begin{equation}
f_{cls}(x,y,\phi_{cls}) \rightarrow p
\end{equation}

\textbf{Phase Two (segmentation):} With the pseudo masks generated in the classification phase, we can form a new training data for the segmentation model as $\mathcal{D}_{seg} = \{(x, p)| x\in \mathcal{X}, p \in \mathcal{P}\}$, where $\mathcal{P}$ is the set of pseudo masks for $\mathcal{X}$. The segmentation model $f_{seg}$ with parameters $\phi_{seg}$ generates the final semantic segmentation results $s$.
\begin{equation}
f_{seg}(x,p,\phi_{seg}) \rightarrow s
\end{equation}

\subsection{Weakly-supervised Pseudo Mask Generation}\label{sec:phase1}
For a giga-pixel whole slide image, defining the presence or absence of the tissue classes in a patch is obviously much easier than carefully drawing pixel-level annotations. So we aim to explore whether patch-level annotations with very limited information are enough for pixel-level semantic segmentation. Zhou~\etal~\cite{zhou2016learning} have demonstrated that classification, localization, detection and segmentation tasks share a similar goal. When training a classification model, the feature maps (Class Activation Maps, CAM) deliver the discriminative object location clues which can be used for object localization and segmentation. Inspired by this, we proposed a novel CAM-based model by first train a classification model. Since the distribution of tissues are somehow random and scatter, it might contain more than one tissue type in one patch. So we define tissue classification as a multi-label classification problem.

\subsubsection{Pseudo Mask Generation}
As demonstrated in Fig.~\ref{fig:network}~(a), given an input patch $x$, we first extract the deep feature maps as follows:
\begin{equation}
f_{cls}(x,\phi_{cls}) \rightarrow m
\end{equation}
where $m$ denotes the extracted feature maps from the last layer.

In order to provide richer and more comprehensive feature representation, we proposed Progressive Dropout Attention (described in Section~\ref{Sec:PD-CAM}) to prevent the classification model from excessively focusing on the most discriminative region.
\begin{equation}
\tilde{m}=\mathcal{A}m
\end{equation}
where $\mathcal{A}$ is the dropout attention map.

After progressive dropout attention, the probability of the $k$-th tissue class $\hat{y}_k$ can be calculated by a global average pooling and a fully connected layer, the same with Zhou~\etal~\cite{zhou2016learning}.
\begin{equation}
\hat{y}_k = \sum \omega_k {\rm GAP}(\tilde{m})
\end{equation}
where ${\rm GAP(\cdot)}$ denotes the global average pooling. Multi-label soft margin loss is applied in the classification network.

With a well-trained multi-label classification model, pixel-level pseudo masks $p$ were generated by Gradient-weighted Class Activation Mapping (Grad-CAM)~\cite{selvaraju2017grad} for the next segmentation model.
\begin{equation}
p = {\rm Grad-CAM}(f_{cls}(x,\phi_{cls}))
\end{equation}

\subsubsection{Progressive Dropout Attention}\label{Sec:PD-CAM}
Although the classification model can provide spatial location hints for the segmentation task. But the goals of these two tasks are still different. As the training process goes further, common classification models tend to focus on the most discriminative part/region of the image, while ignoring some insignificant areas. The activated region shrinkage problem will harm the segmentation task. And it will be amplified in the histopathology image of cancers because the spatial arrangement of different tissue types is relatively random comparing with natural images. Moreover, one-hot encoding labels only contain very limited information. There is still a huge information gap from patch-level labels to pixel-level labels. Therefore, how to maximize the value of such sparse annotations in order to close the gap is still an extreme task. To overcome the above two challenges, we proposed a Progressive Dropout Attention (PDA). Let us start with its basic form, Dropout Attention.

\begin{figure*}[htp]\centering
	\includegraphics[width=.995\linewidth]{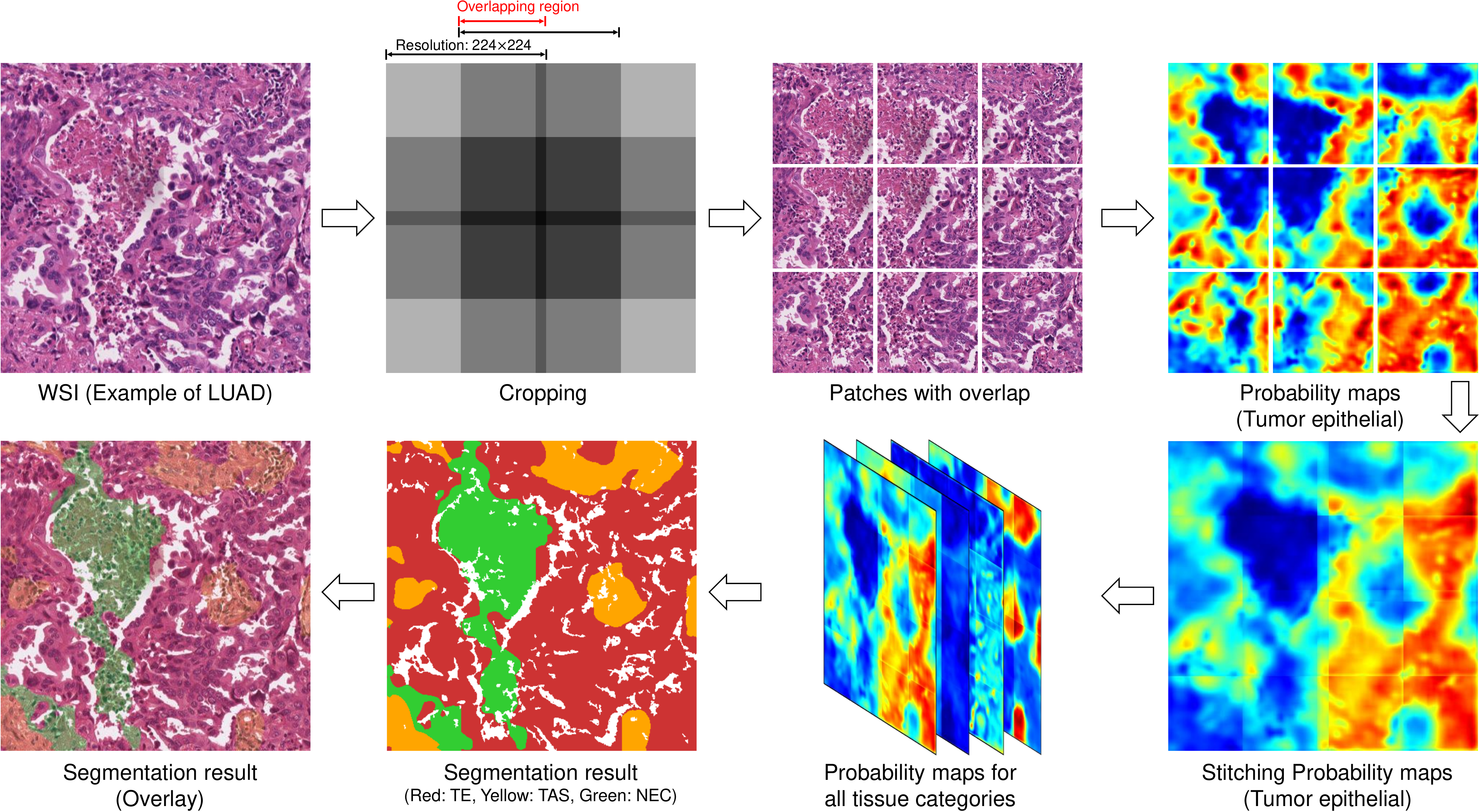}
	\caption{Semantic segmentation for whole slide images. We only show a very small view of the WSI for simple illustration. We show the probability maps of tumor epithelial as the examples.}
	\label{fig:SS-wsi}
\end{figure*}

\textbf{Dropout Attention}:
The idea of the proposed dropout attention is simple and intuitive. We want the neural network to be able to learn as much information as possible from the sparse labels. During the training process, the classification model is not allowed to ``make easy money'' by only relying on the most discriminative areas. On the contrary, the CNN model has to learn more complete and comprehensive spatial information. Therefore, we deactivate the most significant regions in the class activation maps of all the tissue categories, as demonstrated in Fig.~\ref{fig:network}~(a). Such a strategy will weaken the contribution of the most discriminative regions and force the neural network to perform multi-label classification by non-predominant regions, which can effectively expand the activated regions when extracting deep features. According to this idea, we first generate a class activation map (CAM) for each category by the weighted sum of the feature maps $m$.
\begin{equation}
\mathcal{M}_k=\sum\omega_k m
\end{equation}
where $\mathcal{M}_k$ denotes CAM of the $k$-th category.

For each $\mathcal{M}_k$, we set up a dropout cutoff $\beta$ to deactivate the most highlighted area and refresh CAMs as follows.
\begin{equation}
\hat{\mathcal{M}}_k{(i,j)} =
\begin{cases}
\mathcal{M}_k(i,j), & \mathcal{M}_k(i,j) \leq 			\beta\\
0,          & \mathcal{M}_k(i,j) \ \textgreater \beta\\
\end{cases}
\end{equation}
where $i$ and $j$ denote the coordinates, $\hat{\mathcal{M}}$ is the CAM with dropout. Note that, $\beta$ is a relative value which depends on the maximum value of the class activation map.
\begin{equation}
\beta = \mu *\max(\mathcal{M}_k)
\end{equation}
where $\mu$ is the dropout coefficient.

Finally, the dropout attention map $\mathcal{A}$ is the average of all the deactivated CAM.
\begin{equation}
\mathcal{A}(i,j) = \frac{1}{c}\sum_{k=1}^c \mathcal{M}_k(i,j)
\end{equation}

\begin{figure}[htp]\centering
	\setlength{\tabcolsep}{1pt}
	\small
	\begin{tabular}{ccccc}
		\multicolumn{5}{c}{With progressive dropout}\\
		\includegraphics[width=.195\linewidth]{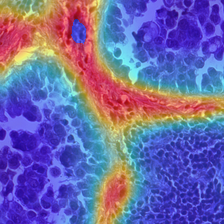}&
		\includegraphics[width=.195\linewidth]{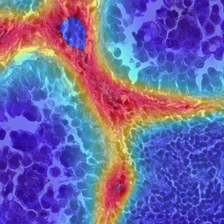}&
		\includegraphics[width=.195\linewidth]{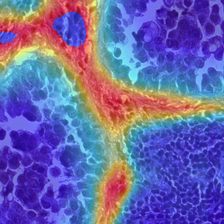}&
		\includegraphics[width=.195\linewidth]{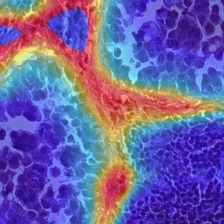}&
		\includegraphics[width=.195\linewidth]{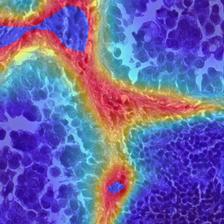}\\
		\multicolumn{5}{c}{Without progressive dropout}\\
		\includegraphics[width=.195\linewidth]{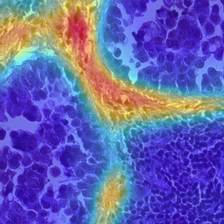}&
		\includegraphics[width=.195\linewidth]{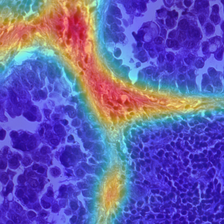}&
		\includegraphics[width=.195\linewidth]{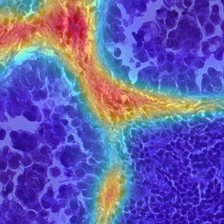}&
		\includegraphics[width=.195\linewidth]{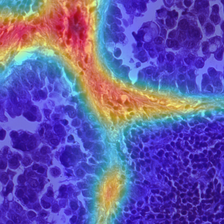}&
		\includegraphics[width=.195\linewidth]{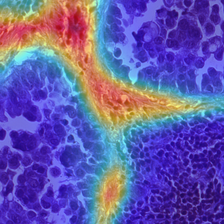}\\
		(a) epoch 3 & (b) epoch 6 & (c) epoch 9 & (d) epoch 12 & (e) epoch 15
	\end{tabular}
	\caption{Examples of progressive dropout attention. We demonstrates the class activation maps $\hat{\mathcal{M}}$ of the tumor-associated stroma regions with (top row) and without (bottom row) dropout in different training epochs. The deactivated areas enlarge with the increasing number of training epochs. (Example from LUAD-HistoSeg)}
	\label{fig:dropout}
\end{figure}

\textbf{Progressive Dropout Attention:}
As we mentioned above, when the training process goes further, the activated area will progressively shrink into a smaller area. According to this observation, we proposed a reverse operation based on dropout attention, called Progressive Dropout Attention (PDA). PDA progressively enlarges the deactivated areas to fight against such a shrinking problem. We redesign the original dropout coefficient $\mu$ to a progressive dropout coefficient, which is no longer a constant value. The progressive dropout coefficient $\mu$ will adaptively decrease when the training epoch increases until $\mu$ meets the lower bound $l$.
\begin{equation}\label{Eq:DAM_beta}
\mu_{t} =
\begin{cases}
\sigma*\mu_{t-1}, & \mu_{t}>l\\
l,          & \mu_{t}\leq l\\
\end{cases}
\end{equation}
where $t$ is the ongoing epoch and $\sigma$ is the decay rate. We set $\sigma = 0.985$ and $l=0.65$ in practice. The initial $\mu$ is set to $1$ at the first three epochs for a better initiation of the classification model. After the $3$-th epoch, we start the dropout and progressively enlarge the dropout area to gradually increase the difficulty of classification.

With progressive dropout attention, the discriminative region shrinkage problem is greatly alleviated and the classification model can learn much richer and wider feature representation and can generate more precise pseudo masks, as demonstrated in Fig.~\ref{fig:dropout}.

\subsection{Pseudo-supervised Tissue Semantic Segmentation}\label{sec:phase2}
In the segmentation phase, we train a semantic segmentation model $f_{seg}$ under the supervision of the pseudo masks $p$, to get the semantic segmentation result $s$ for the input patch $x$.

\begin{equation}
s = f_{seg}(x,p,\phi_{seg})
\end{equation}

Two specific designs, multi-layer pseudo-supervision and classification gate mechanism, were proposed to further improve the semantic segmentation performance in this phase.

\subsubsection{Multi-Layer Pseudo-Supervision}
Due to the information gap between patch-level labels and pixel-level labels, the spatial information learned from the classification network is still incomplete even with progressive dropout attention. To reduce the gap, we have to bring more information to the segmentation model. Since CNN models learn different levels of semantic features at different stages, we generate multi-layer pseudo masks from three different layers to enrich the information. And then we calculate cross entropy loss between the semantic segmentation results and all the pseudo masks.
\begin{equation}
\label{ep:loss}
\mathcal{L}_{seg} = \lambda_{1}\mathcal{L}_{b4\_3} + \lambda_{2}\mathcal{L}_{b5\_2} + \lambda_{3}\mathcal{L}_{bn7}
\end{equation}
where $\lambda_{i} $ is the hyper-parameter. We set $\lambda_1=0.2$, $\lambda_2=0.2$, $\lambda_3=0.6$ in practice. Note that, multi-layer pseudo masks were upsampled to the original image resolution using bilinear interpolation.

\subsubsection{Classification Gate Mechanism}
Long tail problem is common for medical data, especially for histopathology images. For those non-predominant tissue categories, like necrosis and lymphocyte, they will be dominated by the predominant tissue categories. It is easier to generate unsatisfactory pseudo masks for the non-predominant categories than the predominant categories, which may increase the false positive rate in the segmentation phase.

To overcome the long tail problem and to reduce the false positive rate for the non-predominant categories, we proposed a classification gate mechanism. In our proposed framework, we observed that the confidence of the classification results is generally higher than the segmentation results on the question of whether a tissue category exists in a patch image, especially for the non-predominant categories. Because the classification model was trained by ground truth labels while the segmentation model was trained by pseudo masks.

Based on this observation, we introduce a gate for each output channel. Let $o_k\in \mathbb{R}^{d\times n}$ denote the output probability map of the $k$-th tissue category from the segmentation model, where $n$ is the number of categories, $d$ is the dimension of the probability map. For each category $k$, if the predicted probability $\hat{y}_k$ of tissue category from the classification model is smaller than a threshold $\epsilon$, it means a low existence rate of this category. Then we will ``close the gate''  of the probability map $o_k$ by zeroing it.
\begin{equation}
o_k =
\begin{cases}
0, & \hat{y}_k \leq \epsilon\\
o_k, & \hat{y}_k \ \textgreater \epsilon\\
\end{cases}
\end{equation}

Then the semantic segmentation result can be obtained by an $\argmax$ operation of the probability map $o$. We set $\epsilon=0.1$ in practice.
\begin{equation}
s(i,j) = \argmax{o(i,j)}
\end{equation}
where $(i,j)$ denotes the coordination.

\subsubsection{Semantic Segmentation for WSIs}\label{sec:wsi}
The model we defined above is the patch-level semantic segmentation model. Next, we introduce the way we achieve semantic segmentation for the whole slide images. As demonstrated in Fig.~\ref{fig:SS-wsi}, we first cropped patches from a whole slide image with over 50\% overlapping region. With the segmentation model, $n$ channels probability maps can be generated for each patch. Then we stitched the probability maps to the WSI-level. For the overlapping regions, we calculated mean of the probabilities of each category at every pixel location. Then we can obtain the semantic segmentation result of the whole slide image by an $\argmax$ operation.

\subsection{Implementation and Training Details}
In our experiments, all the convolutional neural networks were implemented in PyTorch. The model was trained on an NVIDIA RTX 2080Ti. ResNet38~\cite{wu2019wider} and DeepLab V3+~\cite{chen2018encoder} were introduced as the classification and segmentation backbones respectively. In the classification phase, the model was pre-trained on ILSVRC 2012 classification dataset~\cite{Ahn_2018_CVPR}. The resolution of the patches is $224\times 224$ and the batch size is set to $20$. The number of training epochs is set to $20$, $40$ for LUAD-HistoSeg and BCSS datasets, respectively. All the patches were transformed by random horizontal and vertical flip with the probability $0.5$. We set a learning rate of $1e-2$ with a polynomial decay policy. In the segmentation phase, the number of training epochs and the learning rate for both datasets were set $20$ and $7e-2$, respectively. There is no restriction of the image resolution in the segmentation phase. Several data augmentation methods were applied, including horizontal and vertical flip, Gaussian blur and normalization. 
\section{Datasets}

\begin{figure*}[htp]\centering
	\includegraphics[width=.95\linewidth]{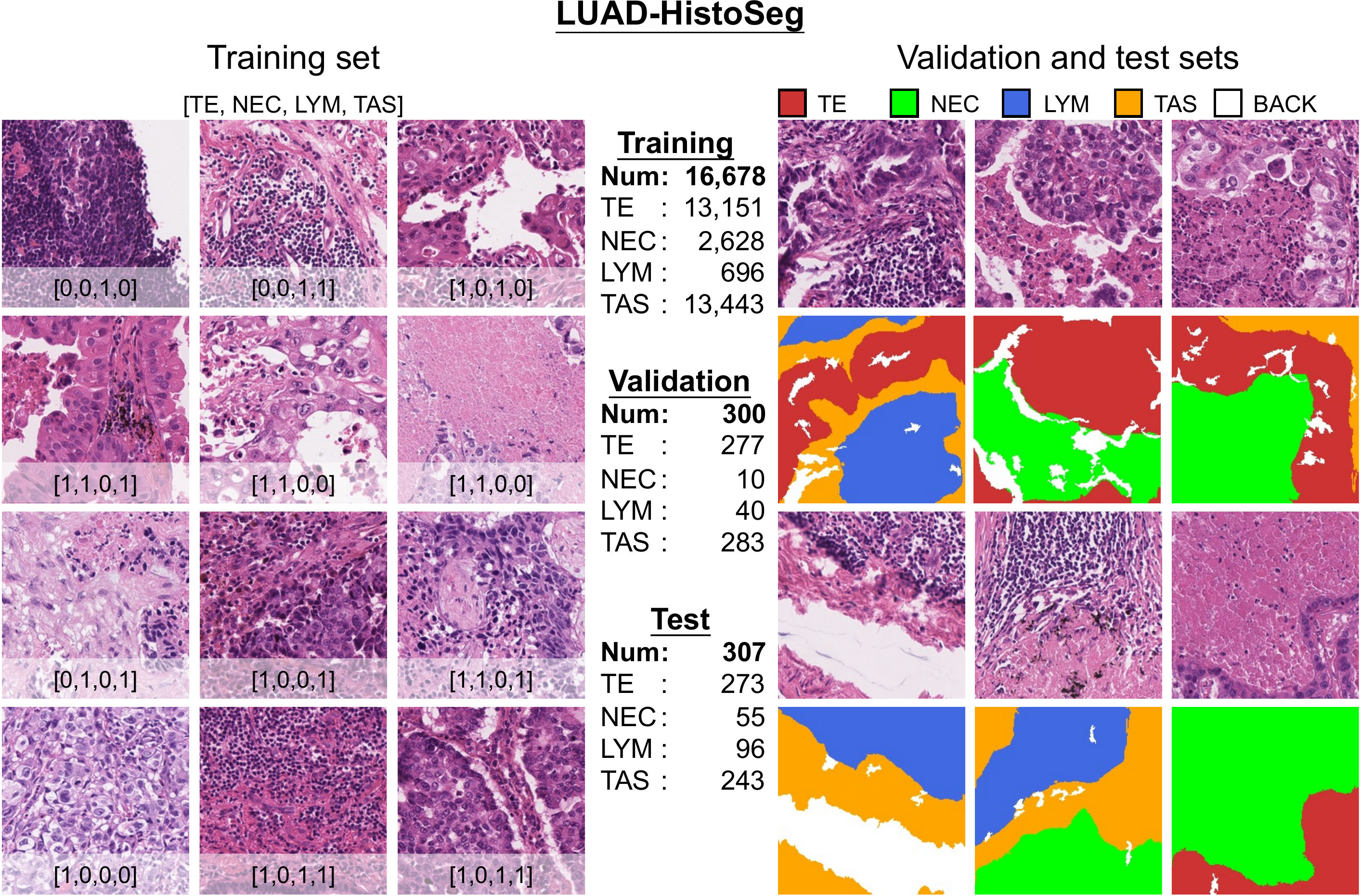}
	\caption{Examples of the released dataset LUAD-HistoSeg. The left-hand side shows the training set with one-hot encoding labels. The right-hand side demonstrates the validation and test sets with semantic segmentation masks. We define four tissue categories in one tissue patch, including tumor epithelial (TE), tumor-associated stroma (TAS), necrosis (NEC) and lymphocyte (LYM). `Num' means the number of patches.}
	\label{fig:luad-histoseg}
\end{figure*}

We evaluate our proposed model on two tissue semantic segmentation datasets, LUAD-HistoSeg and BCSS~\cite{BCSS}.
\subsection{LUAD-HistoSeg Dataset}
As a part of this paper, we release a weakly-supervised tissue semantic segmentation dataset for lung adenocarcinoma, named LUAD-HistoSeg\footnote{Dataset Download Link}, demonstrated in Fig.~\ref{fig:luad-histoseg}. This dataset aims to use only patch-level annotations to achieve pixel-level semantic segmentation for four tissue categories, tumor epithelial (TE), tumor-associated stroma (TAS), necrosis (NEC) and lymphocyte (LYM).

\textbf{Dataset Description: }
29 patients from Guangdong Provincial People's Hospital and 20 patients from TCGA with lung adenocarcinoma were chosen. For each patient, three experienced pathologists (at least ten-year working experience) were asked to examine all the pathology sections and select the most representative section for clinical diagnosis. Each section was scanned by the digital pathology slide scanner (Leica, Aperio-AT2). Then we randomly cropped 800 patches at 10$\times$ objective magnification ($0.0625\mu m/pixel$) with the size of 224$\times$224 for each whole slide image. Next, We dropped the patches with blurry, dirty, large white backgrounds and over-stained problems by a quality control process. We further dropped the ambiguous patches which have classification disagreement among three pathologists. Finally, a total of 17,285 patches were left as our final dataset. We divided them into a training set (16,678 patches, patch-level annotations), a validation set (300 patches, pixel-level annotations) and a test set (307 patches, pixel-level annotations). The data distribution of each category is shown in Fig.~\ref{fig:luad-histoseg}.

\textbf{How to Label: }
We invited five junior clinicians and three experienced pathologists to label all the patches. There are two different kinds of labels, patch-level labels for the training set and pixel-level labels for the validation and test sets. For the training set, annotators have to define whether a specific tissue category is present or absent by a one-hot encoding vector, demonstrated in Fig.~\ref{fig:luad-histoseg} (left). For the validation and test sets, annotators were asked to roughly draw the semantic segmentation masks using Labelme~\cite{labelme2016} and refine the boundaries using PhotoShop, demonstrated in Fig.~\ref{fig:luad-histoseg} (right). Junior clinicians were responsible for labeling and pathologists have to finally confirm the labels. The patches were rejected and dropped if there exists ambiguities. Since lung is mainly composed of the alveolus, there are a lot of white regions randomly distributed in the whole slide image. So we extract these white regions by a color thresholding method. The white backgrounds inside the alveolus were excluded when calculating the performance in all the experiments.

\begin{figure*}[htp]\centering
	\includegraphics[width=.95\linewidth]{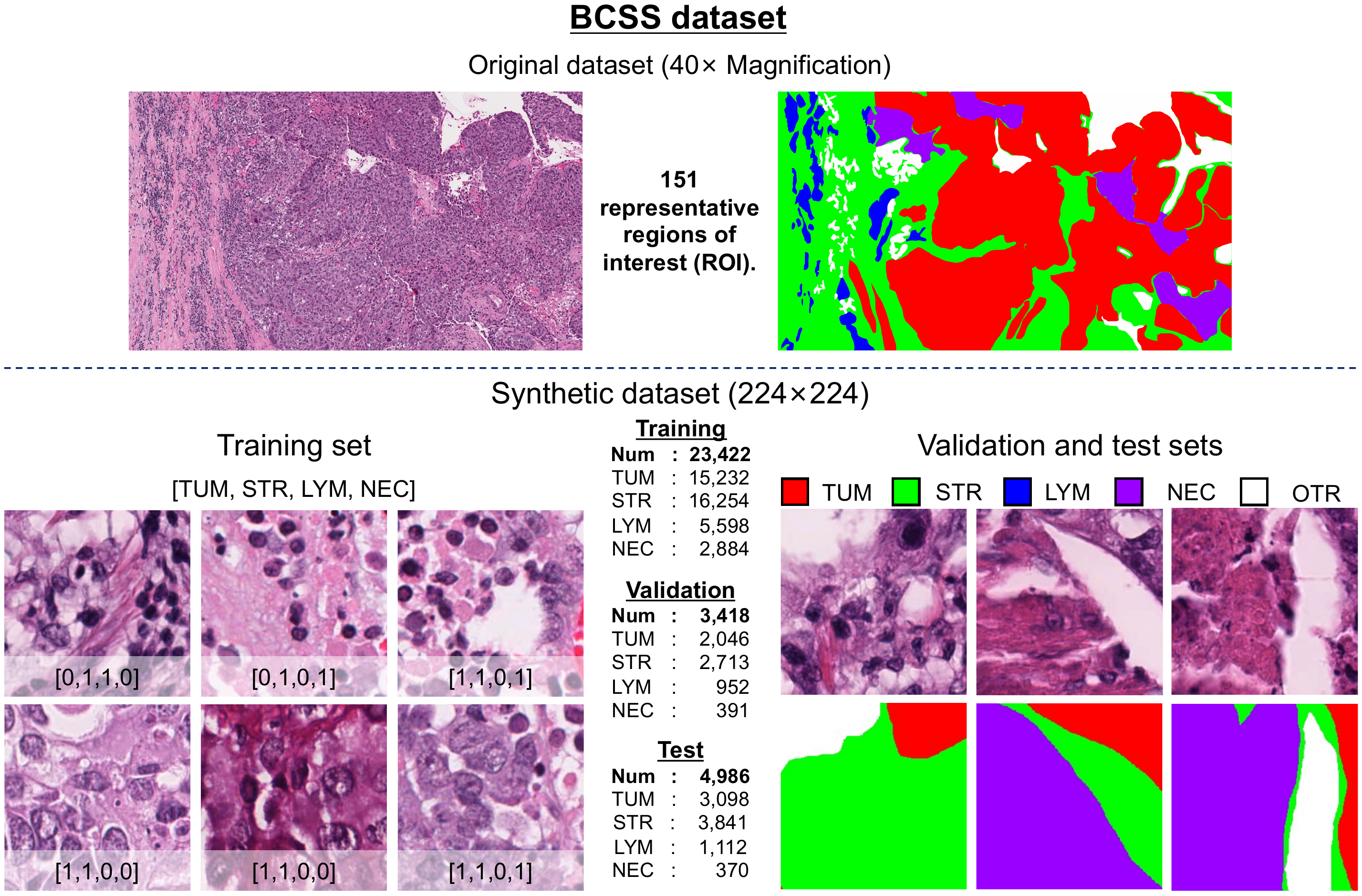}
	\caption{Examples of BCSS dataset. The original BCSS dataset contains 151 large ROIs with pixel-level annotations of five tissue categories, Tumor~(TUM), Stroma~(STR), Lymphocytic infiltrate~(LYM), Necrosis~(NEC) and Other~(OTR). We generate a synthetic dataset for our weakly-supervised approach. The left-hand side shows the training set with one-hot encoding labels. The right-hand side demonstrates the validation and test sets with semantic segmentation masks. `Num' means the number of patches.}
	\label{fig:BCSS}
\end{figure*}

\subsection{Breast Cancer Semantic Segmentation (BCSS) Dataset}\label{sec:BCSS}
We also evaluate our proposed model on a fully-supervised semantic segmentation dataset, to compare our weakly-supervised approach with the fully-supervised approach in order to observe the potential of our proposed model.

Breast cancer semantic segmentation (BCSS) dataset~\cite{BCSS} consists of 151 representative regions of interest (ROIs) from 151 H\&E stained whole slide images of breast cancer, which were selected by a study coordinator, a clinician, and approved by a senior pathologist. The mean size of ROIs is 1.18 $mm^2$ at 0.25 microns per pixel resolution. As shown in Fig.~\ref{fig:BCSS}, the original BCSS dataset provides pixel-level annotations for each ROI with 5 classes, including Tumor (TUM), Stroma (STR), Lymphocytic infiltrate (LYM), Necrosis (NEC) and Other (OTR).

In order to perform weakly-supervised semantic segmentation, we randomly cropped patches from the ROIs and used the semantic segmentation masks to generate one-hot encoding vectors. A total of 31,826 patches were generated and split into a training set (23,422 patches, patch-level annotations), a validation set (3,418 patches, pixel-level annotations), and a test set (4,986 patches, pixel-level annotations), as demonstrated in Fig.~\ref{fig:BCSS}. We also provide the generated patch-level dataset of BCSS via this link\footnote{Dataset Download Link}. 
\section{Experiments}
In this section, we conduct several experiments to comprehensively evaluate the capacity of our proposed model on how well it achieves semantic segmentation using only patch-level annotations. Sec.~\ref{sec:exp-compare} demonstrates the quantitative and qualitative comparisons with state-of-the-art methods. We conduct ablation studies in Sec.~\ref{sec:exp-ablation} to evaluate the effectiveness of our proposed progressive dropout attention, multi-layer pseudo-supervision and classification gate mechanism. Next, we demonstrate the semantic segmentation results of the whole slide images in Sec.~\ref{sec:exp-wsi}. We also measure how much labeling time we can save for the pathologists in Sec.~\ref{sec:manual-vs-model}. We discuss the limitations of the proposed model in Sec.~\ref{sec:exp-limitation}.

We evaluate our proposed model by the following metrics, IoU for each category, Mean IoU (MIoU), Frequency weighted IoU (FwIoU) and pixel-level accuracy (ACC).
\begin{figure*}[t]\centering
	\includegraphics[width=1\linewidth]{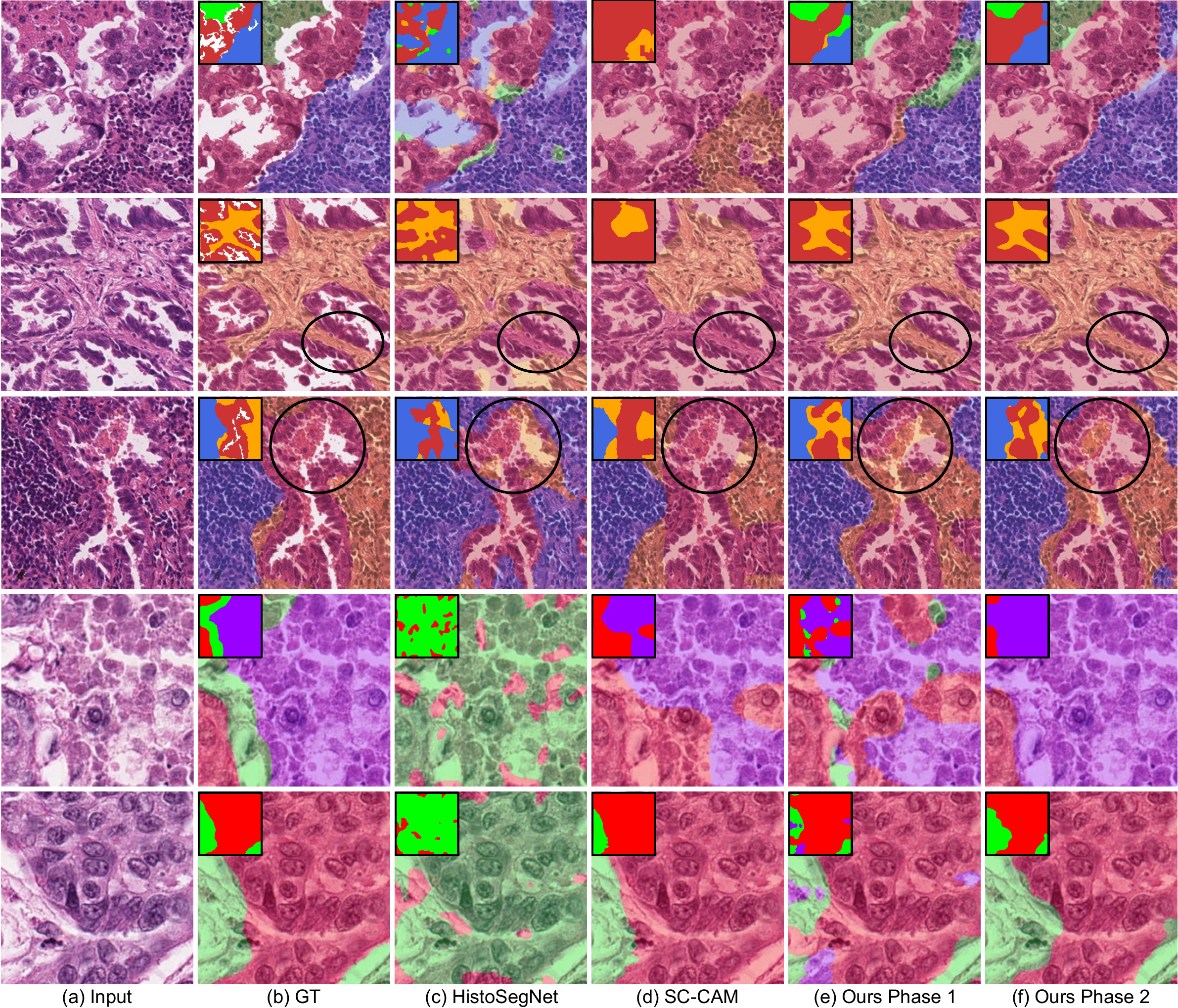}
	\caption{Qualitative results of patch-level semantic segmentation. Results were overlaid on the input images. The upper three rows are LUAD-HistoSeg. The bottom two rows are BCSS. Top-left corner are the semantic segmentation masks. Black circles highlight the inaccurate boundaries and incomplete results of the existing methods.}
	\label{fig:qualitative}
\end{figure*}
\begin{table*}[t]
	\centering
	\caption{Quantitative comparison with existing methods.}
	\begin{threeparttable}
		\begin{tabular}{p{0.11\textwidth}>{\centering}
				p{0.039\textwidth}>{\centering}
				p{0.039\textwidth}>{\centering}
				p{0.039\textwidth}>{\centering}
				p{0.039\textwidth}>{\centering}
				p{0.039\textwidth}>{\centering}
				p{0.039\textwidth}>{\centering}
				p{0.039\textwidth}>{\centering}
				p{0.039\textwidth}>{\centering}
				p{0.039\textwidth}>{\centering}
				p{0.039\textwidth}>{\centering}
				p{0.039\textwidth}>{\centering}
				p{0.039\textwidth}>{\centering}
				p{0.039\textwidth}>{\centering}
				p{0.039\textwidth}}
			\toprule
			\multirow{2}{*}{Method} & \multicolumn{7}{c}{LUAD-HistoSeg} & \multicolumn{7}{c}{BCSS}\cr
			\cmidrule(lr){2-8} \cmidrule(lr){9-15}
			&TE &NEC &LYM &TAS &FwIoU &MIoU &ACC
            &TUM &STR &LYM &NEC &FwIoU &MIoU &ACC \cr
			\midrule
			HistoSegNet\cite{chan2019histosegnet} 						
			&0.45594&0.36302&0.58283&0.50818&0.48538&0.47749&0.65971
			&0.33141&0.46457&0.29047&0.01908&0.37191&0.27638&0.56410\cr
			SC-CAM\cite{Chang_2020_CVPR} 								
			&0.68286&0.64284&0.62063&0.61785&0.64743&0.64104&0.78690
			&0.76788&0.70606&\textbf{0.58023}&0.60073&0.71581&0.66373&0.83427\cr
			\midrule
			Ours Phase 1					&0.75567&0.78079&\textbf{0.73694}&0.69690&0.73324&0.74258 &0.84508 &0.72976&0.68134&0.56191&0.55989&0.68532&0.63323&0.81216 \cr
			Ours Phase 2					&\textbf{0.77704}&\textbf{0.79321}&{0.73406}&\textbf{0.71980}&\textbf{0.75126}&\textbf{0.75603} &\textbf{0.85701} &\textbf{0.78839}&\textbf{0.73157}&0.57295&\textbf{0.66389}&\textbf{0.73745}&\textbf{0.68920}&\textbf{0.84832}\cr
			\bottomrule
			\hspace{1mm}
		\end{tabular}
	\end{threeparttable}
	\label{tab:quantitative}
\end{table*}

\subsection{Quantitative and Qualitative Comparisons}\label{sec:exp-compare}
Table~\ref{tab:quantitative} demonstrates the quantitative comparisons with existing methods. We compare our proposed model with two SOTA CAM-based weakly-supervised semantic segmentation models, one for histopathology images (HistoSegNet~\cite{chan2019histosegnet}) and the other for natural images (SC-CAM~\cite{Chang_2020_CVPR}). We implemented these two papers exactly follow the technical details of the original papers. ``Ours Phase 1'' is the classification model trained in phase one. The semantic segmentation results of this model were generated by Grad-CAM from layer $bn7$. ``Ours Phase 2'' is the semantic segmentation model trained in phase two, which is our final model. As shown in Table~\ref{tab:quantitative}, our final model greatly outperforms both existing models on two datasets.
In LUAD-HistoSeg dataset, even the pseudo masks generated from the classification model in phase 1 can outperform two existing CAM-based WSSS methods in both datasets, which justifies the superiority of our proposed progressive dropout attention. After training the segmentation model in phase 2, our model achieves a significant and consistent improvement in all the categories except LYM in LUAD-HistoSeg. Because LYM only occupies around 4\% in this dataset, which is extremely imbalanced. The lack of training samples may lead to unstable performance. Fig.~\ref{fig:qualitative} demonstrates the qualitative results of different models in both datasets. Our proposed model can generate more precise tissue boundaries comparing with two existing works. HistoSegNet~\cite{chan2019histosegnet} mostly relies on post-processing step to merge the fragile segments. Therefore, it fails to predict complete and unbroken results. Since the distribution of different tissues is relatively random and scatter, but natural images follow some rules like `humans and cars mostly appear on the road'. As highlighted in the black circles, SC-CAM also fails to generate precise boundaries, especially for the non-predominant categories, LYM and NEC. Our proposed progressive dropout attention will deactivate the most discriminative regions and push the neural network to learn more comprehensive features from the entire image. Such design greatly benefits weakly-supervised semantic segmentation in histopathology images. Qualitative results also show that training a segmentation phase of pseudo-supervision is necessary since it can avoid some noisy prediction results in phase one.

\begin{figure}
	\centering
	\includegraphics[width=.985\linewidth]{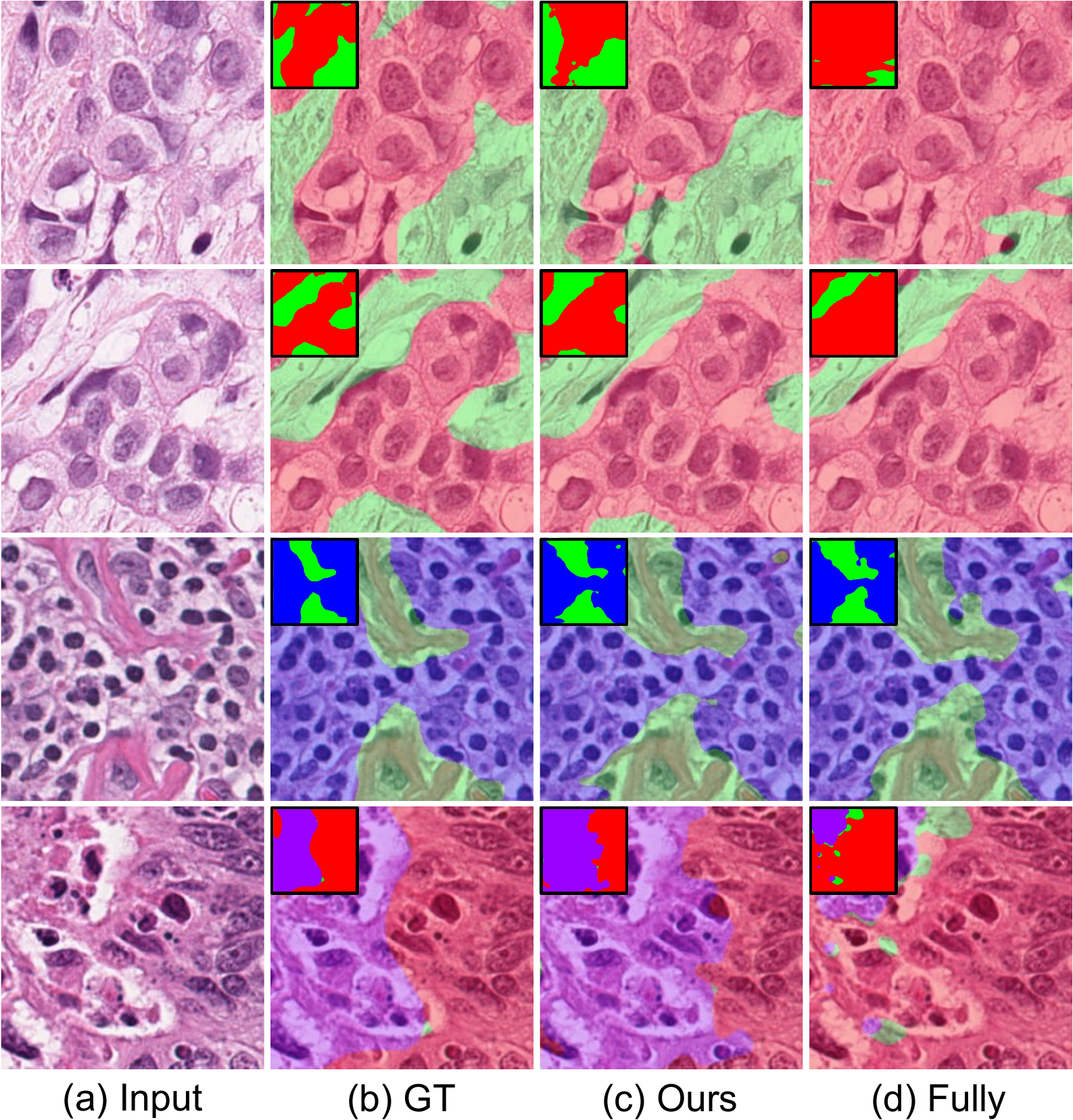}
	\caption{Comparison with fully-supervision (BCSS). Results were overlaid on the input images. Top-left corner are the semantic segmentation masks.}
	\label{fig:qualitative-fully}
\end{figure}

Since BCSS is the tissue semantic segmentation dataset with pixel-level annotations. Therefore, we conducted an additional experiment to evaluate the potential of our proposed model by comparing the proposed pseudo-supervision with fully-supervision. We generated a WSSS dataset from the original BCSS dataset for our proposed model as demonstrated in Sec.~\ref{sec:BCSS}. To be fair, both fully-supervised and pseudo-supervised models were trained on the same network structure DeepLab V3+ with the same training epochs. Comparing with the results generated by the fully-supervised model, shown in Table~\ref{tab:quantitative-fully}, our proposed pseudo-supervised model demonstrates competitive performance for all the tissue categories, even for the non-predominant ones. The performance gap between the pseudo-supervised model and the fully-supervised model is less than 2\%. Fig.~\ref{fig:qualitative-fully} further demonstrates the qualitative comparisons between the pseudo-supervised model with the fully-supervised model. The semantic segmentation results generated by the pseudo-supervised model show visually no difference from the ones generated by the fully-supervised model. Both two models can generate high concordance semantic segmentation results comparing with manual annotations. Unfortunately, when the borders between two tissue categories are not visually clear enough, both two models fail to generate smooth boundaries. It is still a debate whether a smooth and ``accuracy'' boundary is really meaningful for clinical cancer research. Overall, this experiment proves that only relying on patch-level annotations can also achieve superior semantic segmentation results which is good news for pathologists to reduce the annotation efforts.

\begin{table}[t]
	\centering
	\caption{Quantitative comparison with fully supervision.}
	\begin{tabular}{p{0.02\textwidth}>{\centering}
			p{0.041\textwidth}>{\centering}
			p{0.041\textwidth}>{\centering}
			p{0.041\textwidth}>{\centering}
			p{0.041\textwidth}>{\centering}
			p{0.041\textwidth}>{\centering}
			p{0.041\textwidth}>{\centering}
			p{0.041\textwidth}}
		\hline	
		&TUM &STR &LYM &NEC &FwIoU &MIoU &ACC \cr
		\hline
		Ours &{0.78839} &{0.73157} &{0.57295}&\textbf{0.66389}&{0.73745}&\textbf{0.68920}&0.84832\cr
		Fully &\textbf{0.81072} &\textbf{0.74861} &\textbf{0.58680}&{0.59873}&\textbf{0.75310}&{0.68622}&\textbf{0.85760}\cr
		\hline
	\end{tabular}
	\label{tab:quantitative-fully}
\end{table}

\subsection{Ablation Studies}\label{sec:exp-ablation}
\begin{figure*}[t]\centering
	\includegraphics[width=1\linewidth]{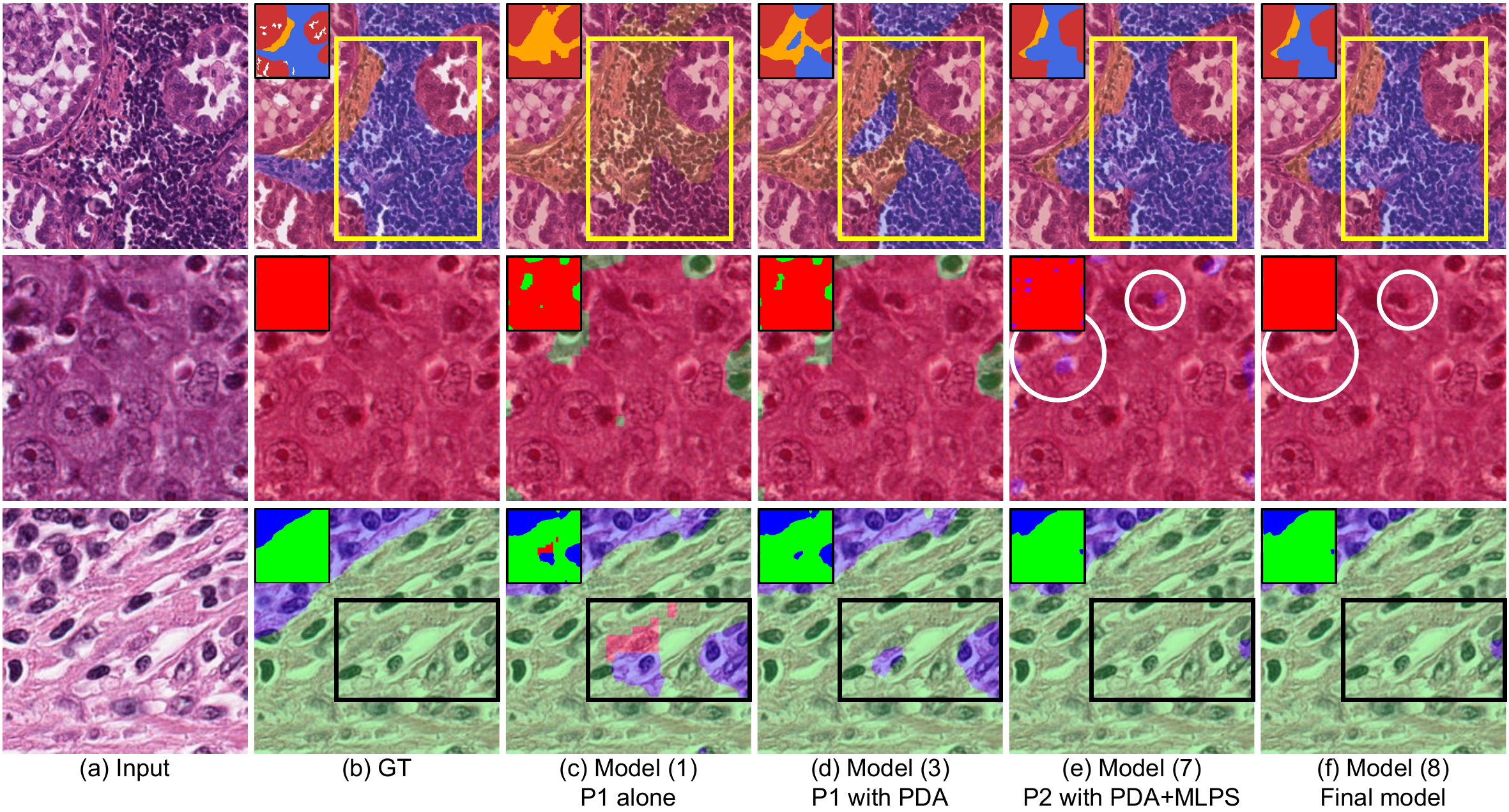}
	\caption{Qualitative results of ablation studies. The first row is from LUAD-HistoSeg. The next two rows are from BCSS. We directly overlaid the results onto the original images. Top-left corner are the semantic segmentation masks.}
	\label{fig:qualitative-ablation}
\end{figure*}
We conducted a series of ablation studies to quantitatively and qualitatively evaluate the superiority of the novelties, including progressive dropout attention (PDA), multi-layer pseudo-supervision (MLPS) and classification gate mechanism. We compared our final model with several baseline models in LUAD-HistoSeg dataset as follows: (1) Phase 1 alone. (2) Phase 1 with dropout attention (DA) with a constant dropout coefficient $\mu=0.7$. (3) Phase 1 with progressive dropout attention (PDA). The results of these three models were the pseudo masks $p_{bn7}$ generated by Grad-CAM from layer $bn7$. (4)-(7) Phase 2 trained by different configurations of multi-layer pseudo masks. (8) Our final model.

The quantitative results are shown in Table~\ref{tab:ablation-quantitative}. We also selected several representative baseline models (1), (3), (7) and (8) to qualitatively prove the effectiveness of the proposed novelties in Fig.~\ref{fig:qualitative-ablation}.

\subsubsection{Progressive Dropout Attention}

\begin{table*}[t]
	\centering
	\caption{Quantitative evaluation: ablation studies. (LUAD-HistoSeg)}
	\begin{threeparttable}
		\begin{tabular}{cccc|cccccccc}
			\toprule
			Phase &PDA &Pseudo Supervision & Class-Gate&TE &NEC &LYM &TAS &FwIoU &MIoU &ACC \cr
			\midrule
			(1) Phase 1 & - & - & -                                     &0.72862&0.72690&0.71305&0.68952&0.71190&0.71452&0.83111\cr
			(2) Phase 1 & DA & - & -                                    &0.75191&0.75568&0.72260&0.69435 &0.72691&0.73113&0.84087\cr
			(3) Phase 1 & $\checkmark$ & - & -                          &0.75567&0.78079&0.73694&0.69690 &0.73324&0.74258 &0.84508\cr
			\midrule
			(4) Phase 2 & $\checkmark$ & $b_{4\_3}$  & -                &0.69942 &0.55688 &0.70002 &0.68347 &0.68295 &0.65995 &0.80978\cr
			(5) Phase 2 & $\checkmark$ & $b_{5\_2}$  & -                &0.75831 &0.77700 &0.67398 &0.67792 &0.71859 &0.72180 &0.83454\cr
			(6) Phase 2 & $\checkmark$ & $bn7$     & -                  &0.77160 &0.74853 &0.72714 &0.70785 &0.74028 &0.73878 &0.84978\cr
			(7) Phase 2 & $\checkmark$ & Multi-Layer & -               &0.77704 &0.78374 &0.73303 &0.71724 &0.74947 &0.75277 &0.85586\cr
			\midrule
			(8) Phase 2 & $\checkmark$ & Multi-Layer & $\checkmark$     &\textbf{0.77704} &\textbf{0.79321}&\textbf{0.73406}&\textbf{0.71980}&\textbf{0.75126}&\textbf{0.75603}&\textbf{0.85701}\cr
			\bottomrule
			\hspace{1mm}
		\end{tabular}
	\end{threeparttable}
	\label{tab:ablation-quantitative}
\end{table*}

In Table~\ref{tab:ablation-quantitative}, model (2) with DA has already achieved an obvious improvement comparing with model (1) in all the tissue categories as well as FwIoU, MIoU and the overall pixel-level accuracy. When equipped with PDA in model (3), the performance continuously improves, especially for the non-predominant categories NEC and LYM. Because deactivating the highlighted areas will push neural networks to learn features from secondary discriminative regions, reducing the information gap between the classification labels and the segmentation labels. But for those non-predominant categories, drastically increasing the difficulty may bring adverse effects. Therefore, progressively increasing the difficulty can smooth the training process, resulting in a better performance improvement. Fig.~\ref{fig:qualitative-ablation}~(c)~\&~(d) shows the results of model (1) and model (3). In the yellow boxes, we can observe the lymphocyte regions from the model with PDA have higher concordance with ground truth comparing with the model without PDA. Although the pseudo masks are still imperfect, we successfully reduce the information gap between image-level labels and pixel-level labels by correcting some false predicted labels.

Besides the improvement of the semantic segmentation performance, we also want to know whether PDA will greatly harm the classification results, which is not our expectation. Table~\ref{tab:classification-PDA} demonstrates the classification results of the classification model after applying PDA. We can find that the overall accuracy only decreases around 1\% in LUAD-HistoSeg and less than 0.1\% in BCSS. We believe that it is worth to trade-off less than 1\% classification accuracy for more than 2\% semantic segmentation improvement.

\begin{table}\centering
	\caption{Classification Results with and without PDA (patch-level accuracy).}
	\begin{tabular}{l|ccccc}
		\hline
		& LUAD-HistoSeg(20-th epoch) & BCSS(30-th epoch)\cr
		\hline
		P1 w/o PDA &0.93893 &0.90784\cr
		P1 w PDA   &0.92508 &0.90694\cr
		\hline
	\end{tabular}
	\label{tab:classification-PDA}
\end{table}

\subsubsection{Multi-Layer Pseudo Supervision}
Since the information gap between patch-level classification labels and pixel-level segmentation labels is huge. The pseudo masks generated from patch-level annotations are no doubt incomplete and imperfect. It is the reason why we proposed MLPS to provide as much information as possible from different layers of the classification model. To evaluate the effectiveness of MLPS, we compare the proposed MLPS model with the model trained by single layer pseudo masks, demonstrated in Table~\ref{tab:ablation-quantitative}. Among the models trained by the pseudo masks from a single layer, model (6) trained by $p_{bn7}$ shows the best performance because it is the closest layer to the inference with the finest semantic information. The model trained by all three layers outperforms all the single layer pseudo masks models. Fig.~\ref{fig:qualitative-ablation} demonstrates the results with and without MLPS. In the yellow and black boxes, results generated by model (7) with MLPS are more complete and achieve higher concordance with ground truth. Experimental results prove that introducing multi-layer pseudo masks can provide more information than the pseudo mark from a single layer. And the incorrect noisy pseudo labels can also be regarded as the regularization method to avoid overfitting.

\subsubsection{Classification Gate Mechanism}
Classification gate mechanism is proposed to reduce the false-positive rate for the non-predominant tissue categories. Model (8) and Model (7) in Table~\ref{tab:ablation-quantitative} demonstrate the models with and without classification gate mechanism in LUAD-HistoSeg, respectively. For the predominant categories, tumor epithelial (TE) and tumor-associated stroma (TAS), classification gate mechanism gets a very slight improvement because the predominant ones occupy more than 60\% of the samples. The segmentation model can learn a better feature representation of them, which results in a lower false-positive rate. For the non-predominant one necrosis (NEC) in LUAD-HistoSeg, classification gate mechanism improves the IoU by more than 1\%. Fig.~\ref{fig:qualitative-ablation}~(f)~\&~(e) demonstrates the results with and without gate. In the white circles, false-positive results have been successfully corrected by the classification gate mechanism.

\begin{figure*}[t]
	\centering
	\includegraphics[width=.99\linewidth]{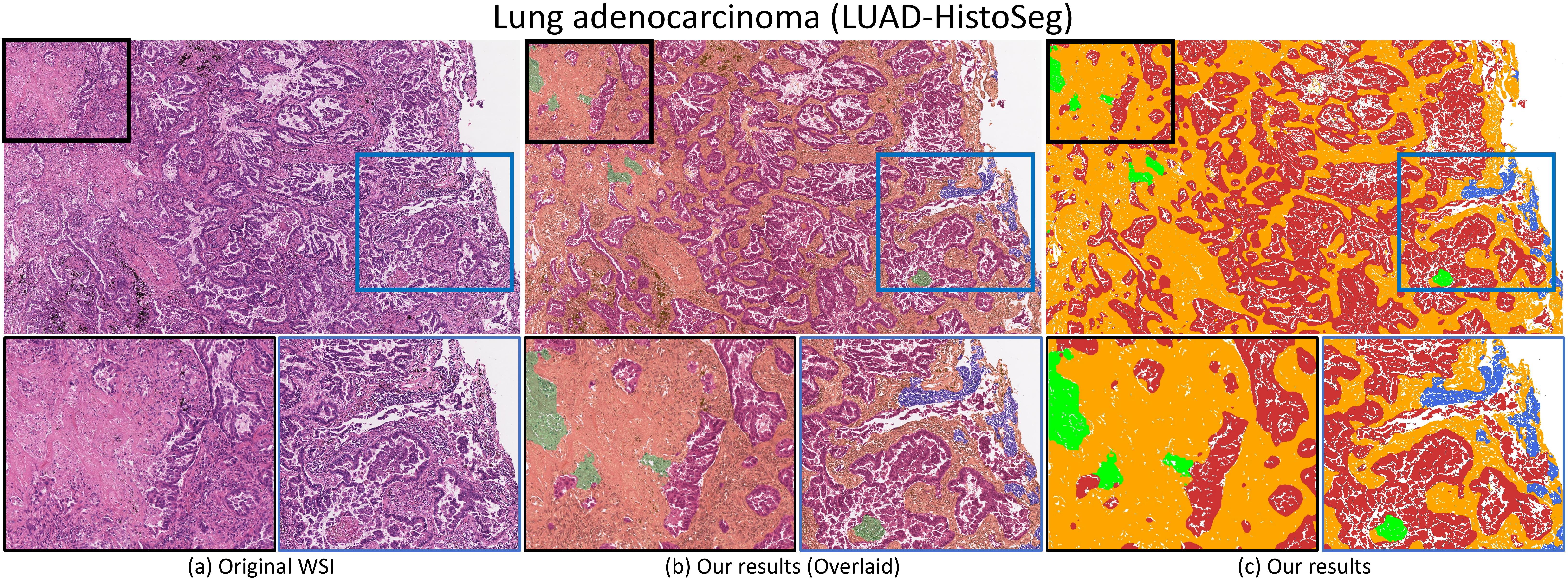}\\
	\includegraphics[width=.99\linewidth]{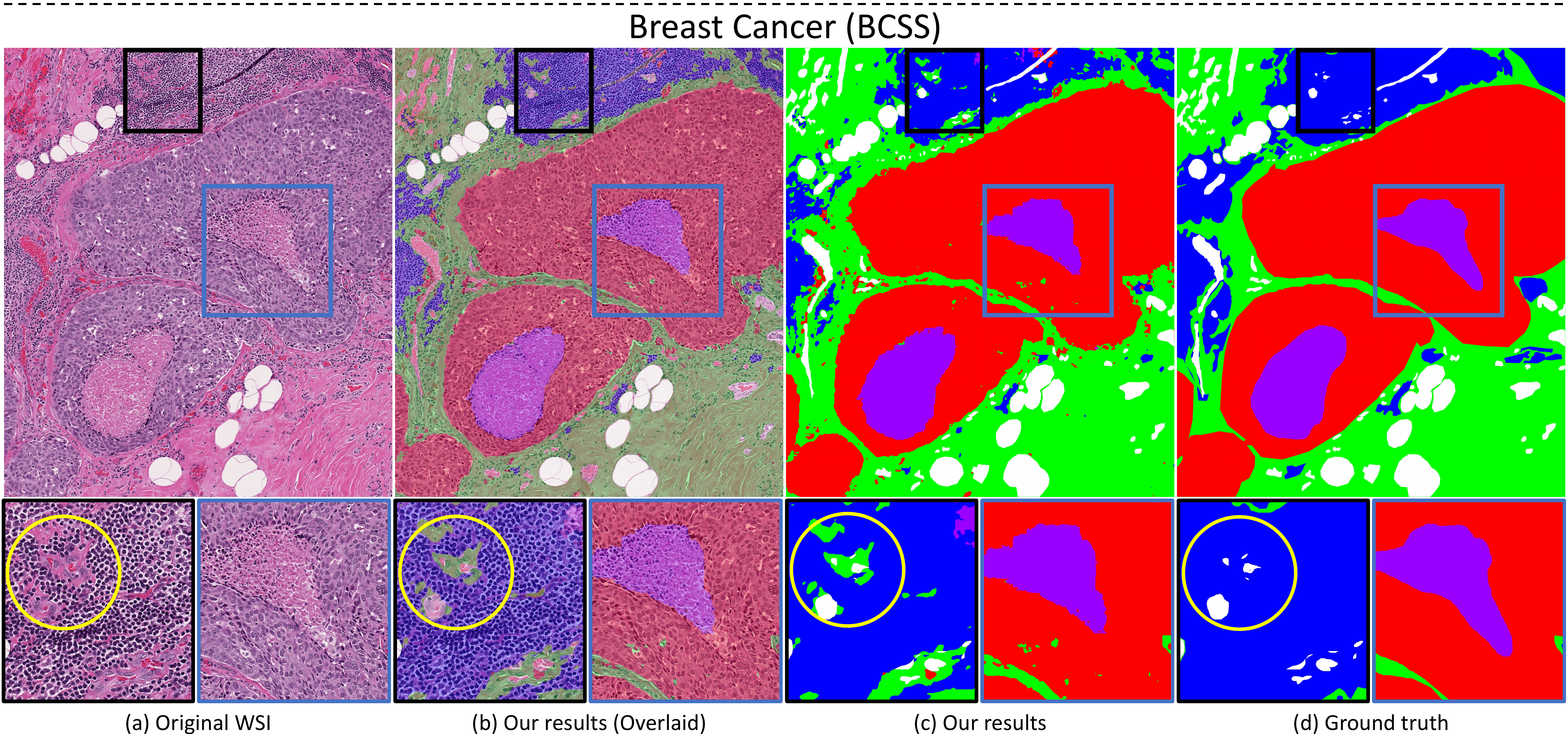}
	\caption{Semantic segmentation results of the whole slide image in lung adenocarcinoma and breast cancer. We show two zoom in regions in the black and blue boxes. Yellow circle shows the ambiguous region. Since the resolution of the WSI is huge, we only demonstrated a small part it.}
	\label{fig:qualitative-wsi}
\end{figure*}

\subsection{Qualitative Results of WSIs}\label{sec:exp-wsi}
In Fig.~\ref{fig:qualitative-wsi}, we also demonstrate the semantic segmentation results of two whole slide images with lung adenocarcinoma and breast cancer, respectively. The way we generate WSI-level semantic segmentation is shown in Sec.~\ref{sec:wsi}. Since BCSS was originally introduced for fully-supervised semantic segmentation, we can compare our results with manual annotations. In both lung adenocarcinoma and breast cancer WSIs, our proposed model can generate visually pleasing results. We can found that the predominant categories such as tumor epithelial and tumor-associated stroma have high concordance comparing with the ground truth labels, while the non-predominant categories necrosis and lymphocyte have lower concordance but are still visually pleasing.

When zooming in the whole slide images (highlighted in black and blue boxes), some ``imperfect'' results can be found such as the unsmooth region boundaries and some very small isolated regions. The reason why we double quote ``imperfect'' is that it is hard to decide whether such results are inaccurate or not. For example, the ROI in the yellow circle, there are some small stroma regions inside the lymphocytic infiltrate region. Globally speaking, they should be categorized as the lymphocytic infiltrate regions but they have the same morphological appearance with stroma. Furthermore, the borders between different tissue types are commonly ambiguous, especially for the tumor invasive regions. It is still a debate whether a smooth and ``accuracy'' boundary is really meaningful for clinical cancer research.

\subsection{How Can We Reduce Annotation Efforts?}\label{sec:manual-vs-model}
We also conducted an experiment to quantitatively evaluate the reduced efforts of manual annotation by applying our proposed model. We randomly selected 100 patches (224$\times$224) from LUAD-HistoSeg dataset. Three junior pathologists were invited to join this test. Pathologists were first asked to label patch-level annotations by our developed tiny tool. For each category, there are two buttons, `$\checkmark$' and `$\times$', to decide whether a tissue category is present or absent. Next, pathologists were asked to use Labelme~\cite{labelme2016} to draw pixel-level annotations. There is no doubt that answering ``Yes or No'' questions is more efficient as shown in Table.~\ref{tab:time}. All three pathologists only spent less than 10 minutes to finish 100 patch-level annotations while average around 200 minutes for pixel-level annotations. We also observe that pathologists often struggled and spent much time refining the boundaries when doing pixel-level annotations, while patch-level annotations can avoid this. Besides time efficiency, patch-level annotations also have higher consistency than pixel-level annotations. We measure the consensus score by dividing the sum of agreeing labels by the total number of labels. The consensus scores of patch-level and pixel-level annotations are 92.25\% and 85.64\%.

\begin{table}[t]
	\centering
	\caption{Comparisons of the timing statistics for patch-level annotations and pixel-level annotations.}
	\begin{tabular}{c|c|c}
		\hline
		Pathologists & Patch-level (minutes) & Pixel-level (minutes)\\
		\hline
		1	& 6.8 & 177.5\\
		2	& 5.2 & 209.3\\
		3	& 7 & 231.6\\
		\hline
	\end{tabular}
	\label{tab:time}
\end{table}

\subsection{Limitations}\label{sec:exp-limitation}
There are still some limitations of our proposed model. It has achieved outstanding performance for the predominant tissue categories. But for the non-predominant ones, lack of enough training samples is always the greatest barrier towards precise segmentation results. Collecting more training samples for these categories may alleviate this problem. Second, as discussed in Sec.~\ref{sec:exp-wsi} (Yellow circle in Fig.~\ref{fig:qualitative-wsi}, our model recognize some small stroma regions inside the lymphocytic infiltrate region. Because the model only considers the morphological features within the receptive field, which may introduce a lot of isolated regions inside a large region. Actually in clinical practice, pathologists define a tissue category by not only observing the morphological appearances locally but also considering a large surrounding area of microenvironment globally. Introducing a global-local design may be a solution to solve this problem and we will keep on discovering it in future works. 
\section{Conclusion}
In this paper, we proposed a tissue-level semantic segmentation model for cancer histopathology images. The major contribution of this model is to replace pixel-level annotations with patch-level annotations, which is significant progress for pathologists to reduce their annotation efforts. Our proposed model achieves competitive performance with fully-supervised model, which means that pathologists only need to define the presence or absence of the tissue categories in a patch instead of carefully drawing the labels. In methodology, we proposed several technical novelties to minimize the information gap between patch-level and pixel-level annotations, and achieved outstanding semantic segmentation performance. To contribute to the research fields of computational pathology and cancer research, we also introduce a new weakly-supervised semantic segmentation dataset for lung adenocarcinoma, LUAD-HistoSeg. This is the first tissue-level semantic segmentation dataset for lung cancer. By applying our proposed model, we also keep on generating more tissue-level semantic segmentation datasets for different cancer types. More senior pathologists will be invited to join this project for labels verification. Hopefully, these datasets will be released soon.

%
\bibliographystyle{IEEEtran}
\bibliography{tpami}
\end{document}